\documentstyle[11pt]{article} 
\input{psfig} 
\begin{document} 
\begin{flushright}  
{BONN-TH-29-P01}\\   
\end{flushright}  
\vskip 2 cm 
\begin{center} 
{\Large {\bf A note on two-loop effects in the DMSSM}} 
\\[0pt] 
\bigskip 
\vspace{0.93cm} 
{\large 
{\bf D.M. Ghilencea$^a,$\footnote{
{{ {\ {\ {\ E-mail: Dumitru@th.physik.uni-bonn.de}}}}}} } 
}\\[0pt] 
\vspace{0.23cm} 
$^a${\it Physikalisches Institut der Universitat Bonn,} \\
{\it Nussallee 12, 53115 Bonn, Germany.}\\
\bigskip 
\bigskip 
\vspace{3.4cm} Abstract
\end{center} 
{\small 
We investigate the proposed ``D-brane alternative'' to the MSSM model (DMSSM)
which is a type II B string orientifold model with N=1 supersymmetry,
three generations and a 
$SU(3)\times SU(2)_R\times SU(2)_L \times U(1)_{B-L}$ gauge group.
An accurate analysis at two-loop level is performed 
to show that unification constraints
 predict a ``left-right'' symmetry breaking scale in
the TeV region. The exact value of this scale is the result of the
competing effects of the two loop terms against the
low energy supersymmetric threshold effects.
The model accommodates  {\it logarithmic}
unification of the gauge couplings 
at an intermediate scale of $10^{12}$ GeV and 
the necessary conditions to achieve this are addressed. 
}
\newpage

\section{Introduction}
The Minimal Supersymmetric Standard Model (MSSM) is currently the 
most studied supersymmetric extension of the Standard Model, and it 
provides a consistent framework for investigating the  phenomenological
aspects and possible signatures of low energy supersymmetry. 
The MSSM model may be regarded as  the low energy limit of four
dimensional heterotic string models where the latter
provide a fully unified theory with gravity and may predict upon 
appropriate compactifications the MSSM  gauge group and
massless spectrum  and amount of supersymmetry. 
Examples in this respect are provided by the $E_8\times E_8$ 
models compactified on a Calabi-Yau manifold \cite{Candelas:1985en}
leading to the $E_6$ gauge group which may then  be broken in the
presence of Wilson lines \cite{Witten:1985xc} to a 
Standard Model-like gauge group. Further, 4D N=1 supersymmetry is 
broken by non-perturbative effects \cite{Derendinger:1985kk}.
In a compactified string theory standard
$SU(5)$-like hypercharge normalisation as in the MSSM is 
possible without a stage of grand unified group, as this
depends on the Kac-Moody level used. For level-1 case an $SU(5)$
relationship emerges, even though the group is just the Standard Model.
This is one of the  possibilities as other 
$U(1)_Y$  normalisations are allowed  
(for related developments see \cite{Dienes:1996sq}). Nevertheless
such $SU(5)$-like normalisation together with the massless spectrum
and symmetry group of the
MSSM enable one to claim circumstantial evidence for (a logarithmic)
unification of the gauge couplings within this model or similar ones. 
Further, the unification is in general stable \cite{Ghilencea:1998mu}  
under the inclusion (in addition to the MSSM spectrum) of extra 
heavy states predicted by the string. Indeed, taking as input the 
low energy values of $sin^2\theta_{W}(M_Z)$ 
and $\alpha_3(M_Z)$ and
performing a two-loop RG flow to include radiative effects, leads to 
a value of the unification scale equal to $\approx 2\times 10^{16}$
GeV. This value  is within a factor of 20 \cite{Dienes:1997du} (see also
\cite{Dienes:1996sq}) from  the heterotic string  prediction
obtained after fixing the string tension to the value leading to the correct
gravitational strength coupling. 
Such discrepancy factor may be regarded as a mismatch of the two scales,
but  may not be so significant given the many orders of magnitude 
over which we extrapolate the running of the gauge couplings. 
Further, this discrepancy may be accounted for by string threshold 
corrections in the presence of Wilson line background \cite{Nilles:1996kb}
without going to the strong coupling regime of the string.
Therefore the unification of the gauge couplings in MSSM-like
models may be regarded as circumstantial support for 
supersymmetry and a fully embedding of the MSSM 
into the (weakly coupled) heterotic string theory may be possible.

Alternative possibilities exist in the context of strongly coupled
heterotic scenarios (M theory). 
In this case the string scale \cite{Witten} $M_{s}\sim g\,M_{P}\,e^{-\phi }$
where $g$ is the gauge coupling, $\phi $ is the dilaton, $M_{P}$ is the
Planck mass. Such a relation allows for a low string scale through the
choice of the dilaton v.e.v. $<\phi >$. This has caused much interest for
it may bring \cite{Witten} the  string scale prediction into better 
agreement with the aforementioned MSSM  unification scale. 
It was also noticed \cite{Lykken:1996fj}
that the mechanism can  be applied to lower the string scale even
further, perhaps even down to the ``TeV region'', to give a  low 
compactification scale \cite{Dienes:1999vg}, \cite{Arkani-Hamed:1998rs}
and a low string scale as well. 

With growing interest in the physics of large extra dimensions and 
low scale string models,
alternative, low energy supersymmetric models to the MSSM
were suggested \cite{Dienes:1999vg}, \cite{Aldazabal:2000sa},
\cite{Aldazabal:2000tw}, \cite{Aldazabal:2000sk}. 
The presence of a low string scale may require a significant 
change of gauge couplings running, if the couplings are still supposed
to meet at the
string scale. In some cases this may be explained by threshold effects  
power-like in the scale \cite{Dienes:1999vg},\cite{Antoniadis:1999ge}, due
to Kaluza Klein states, although these seem to bring some amount of
fine-tuning  \cite{Ghilencea:1998st}.
This may in principle be avoided if the couplings unify not at the string
scale but at the first winding mode above it \cite{Antoniadis:1999ge}
\cite{Bachas:1998kr}, which may be close to
the Planck scale, restoring  MSSM-like logarithmic unification.
However, consistent model building along these ideas lacks
the accuracy and consistency of the situation 
MSSM vs. heterotic string case and in general, low (string) scale unification
(power-like  or not) may not be easy achieve \cite{Ghilencea:2000dg}. 

One of the possible solutions to deriving low energy supersymmetric
models which bear some similarities to the MSSM case, but have a {\it
  low} unification/string scale, was provided by 
type IIB $Z_N$ orientifold models \cite{Aldazabal:2000sa}, 
\cite{Aldazabal:2000sk} with D3 branes placed at $C^3/Z_N$
singularities.
These models reproduce the desirable features of a particle theory
model: they have N=1 supersymmetry,  SM gauge group (at least below
some scale), three quark-lepton generations. One characteristic they 
bring is  a non-standard hypercharge normalisation. 
Examples of this type have been analysed at 
one loop level \cite{Aldazabal:2000tw}, \cite{Aldazabal:2000sk} and
 at two loop level \cite{Ghilencea:2000dg}.
The purpose of this work is to investigate another model of this
class  \cite{Aldazabal:2000sk} the so-called  ``D-brane
alternative'' to the MSSM (DMSSM). A one loop analysis
\cite{Aldazabal:2000sk} has shown
that the model presents a logarithmic unification of the couplings at
a scale close to $10^{12}$ GeV and may be able to fit the experimental 
constraints on $\alpha_3(M_Z)$, $\sin^2\theta_{W}(M_Z)$ and 
$\alpha_{em}(M_Z)$. The reason why the couplings may unify
at a {\it lower} (than in the MSSM) 
scale is not due to  power-like
running, but to the different symmetry group and spectrum above some
 ``left-right symmetry'' breaking  scale $M_R$. We argue that 
for an accurate investigation
one should perform a two loop analysis, given the present
accuracy of low energy experimental data. We employ a simple method to
perform such a two loop investigation and analyse 
the constraints this model must respect in order to achieve 
logarithmic unification. In particular we  stress the importance of competing 
effects of the one-loop low energy supersymmetric
thresholds\footnote{Such one-loop threshold effects 
are comparable to two loop effects.} against pure two-loop effects. 
These effects have strong implications for
the existence  of a ``left-right'' symmetry  breaking scale 
$M_R$, whose  value and correlation to one-loop
supersymmetric thresholds is  analysed.

\section{Description of the DMSSM}
A brief outline of the DMSSM model and its rather distinct features
relative to the MSSM case are outlined below.
\begin{itemize}
\item{The gauge group above the scale $M_R$ is a minimal 
``left-right'' extension of the Standard Model,
$SU(3)\times SU(2)_L\times SU(2)_R\times U(1)_{B-L}$. 
Below the scale $M_R$ the usual $SU(3)\times SU(2)_L\times U(1)_Y$
gauge group applies.}

\item{The charge of the $U(1)_{B-L}$ group as
well as that of the hypercharge group $U(1)_Y$ (below $M_R$) 
have non-standard normalisation, $k_{B-L}=32/3$ and $k_y=11/3$.}

\item{The unification scale is lowered from the MSSM case. This is not the 
result of power-like RG flow of the couplings, but of the 
non-standard  normalisation of the $U(1)$ groups and the enhanced
gauge symmetry (and also different spectrum above $M_R$). 
This provides a specific  model with
low scale {\it logarithmic} unification.}

\item{The representations predicted by this model have the following 
quantum numbers  with respect to the aforementioned 
 gauge group above $M_R$: the Higgs
sector: $3\times(1,2,2,0)$,
the quarks sector: $3\times (3,2,1,1/3)+3\times(\bar 3,1,2,-1/3)$,
the leptons sector: $3\times (1,2,1,-1)+3\times(\bar 1,1,2,1)$.
The model has the nice feature of  predicting three generations
as a result of three additional complex (compact) dimensions 
\cite{Aldazabal:2000sk}.}

\end{itemize}
As shown, the gauge group above the scale $M_R$ is enhanced from
that of the MSSM. One  important
consequence  is the non-standard  normalisation of the
charges of the $U(1)$ above and below the scale $M_R$
as discussed below. 
The initial starting gauge group in the DMSSM
contains $U(3)\times U(2)_L\times U(2)_R$
which includes three $U(1)$ gauge groups. Of these $U(1)$'s only one is
anomaly-free, with the other two (anomalous) $U(1)$ to become massive
and decouple due to a generalised  Green-Schwarz mechanism
\cite{Aldazabal:2000tw}, \cite{Aldazabal:1998mr}. As a
consequence of the presence of three independent $U(1)$'s 
coming from the non-Abelian sector (rather than two,
as in the MSSM), the hypercharge normalisation will be different
since the number of non-Abelian gauge groups controls the normalisation of
the  anomaly-free $U(1)$ group \cite{Aldazabal:2000sa} 
as given by the formula $k_y=5/3+2(N-2)$ where $N$ is
the number of Abelian U(1)'s, each coming from a non-Abelian group
of the model \cite{Aldazabal:2000sa}. Thus for a starting non-Abelian
gauge group $U(3)\times U(2)$ as in the MSSM we have $N=2$ and $k_y=5/3$
emerges. For a starting non-Abelian  group
 $U(3)\times U(2)_L\times U(2)_R$ as in the DMSSM,
the same formula gives $k_y=11/3$ ($N=3$). 
For the same model one also  has
$k_{B-L}=8/3+8(N-2)$, after using the relations 
between hypercharge, $SU(2)_R$ and $U(1)_{B-L}$ generators.
This gives $k_{B-L}=32/3$,  $(N=3)$ which is different
from the standard $SO(10)$ embedding corresponding to $k_{B-L}=8/3$.
This observation has strong implications for the boundary
conditions of the running of the associated gauge couplings and the
value of the unification scale. Further,
these boundary conditions are exactly those derived from embedding the
gauge group in a D-brane scheme \cite{Aldazabal:2000sk}
with D-branes placed at $C^3/Z_N$, $(N=3)$ singularities. 
Note that $N=3$ case ``fixing'' the normalisation of the 
anomaly-free $U(1)$'s 
and related to the (number of) non-Abelian gauge groups to start 
with, is also related to the number of generations (three) and 
of complex 
dimensions equally twisted  \cite{Aldazabal:2000sa}.

Since the  gauge group above the scale $M_R$ is enhanced with 
states also charged under $SU(2)_R$ in addition to 
$SU(3)\times SU(2)_L$,  
the running of the gauge couplings above the scale $M_R$ will be 
affected significantly. If compared to the MSSM,
two-loop effects will be enhanced, and for an equally accurate
analysis in the DMSSM, a two loop evaluation of the RG flow is necessary.
This is  due to the fact that for rather similar
matter spectrum, the wavefunction
renormalisation of the matter fields (two loop effect for gauge
couplings) also receives gauge corrections from the additional
gauge bosons of the $SU(2)_R$ group. Such enhancement of the two
loop effects is rather generic in models with larger gauge group.
Further, a simple one-loop result for the DMSSM shows that
\begin{equation}\label{eq1}
\alpha_3^{-1}(M_Z)=-\frac{15}{2\pi} \ln\frac{M_R}{M_Z}+
\frac{3}{2}\left(1-4\sin^2\theta_{W} \right)\alpha_{em}^{-1}(M_Z)
-\frac{15}{2 \pi}\ln \frac{T_{eff}}{M_Z}
+two-loop
\end{equation}
where $T_{eff}$ accounting for the low energy supersymmetric 
threshold effects will be defined later.
One important observation of this one loop result
in case $T_{eff}\approx M_Z$,   is that the 
value of the scale $M_R$ (parameter of the model) 
is required to have values relatively close to the electroweak 
scale $M_Z$ due to  the requirement $\alpha_3(M_Z)$ be positive, 
given  the experimental value of
$\sin^2\theta_{W} (M_Z)|_{0}=0.23114\pm 0.00016$ \cite{pdg}.  
These requirements  may change upon including two loop
terms and it is thus important to know how stable these one loop
bounds on $M_R$  are. Further, these bounds are affected by the 
effects of $T_{eff}$ which are also comparable to two loop effects.
This  issue is relevant because the value of $M_R$ 
should not be too large, otherwise one would have to explain why
the mass of the two Higgs pairs  (whose mass we set for simplicity 
to $M_R$)  present in addition to the MSSM Higgs sector
should be large compared to that of the usual Higgs. 

In the case  of the MSSM a similar calculation gives
\begin{equation}\label{eq2}
\alpha_3^{-1}(M_Z)=
\frac{1}{7}\left(15 \sin^2\theta_{W}-3 \right)\alpha_{em}^{-1}(M_Z)
+\frac{19}{28 \pi}\ln \frac{M_{eff}}{M_Z}+two-loop
\end{equation}
with $M_{eff}$ to account for low energy supersymmetric thresholds
\cite{Carena:1993ag} different from the DMSSM case.
Comparing equations (\ref{eq1}), (\ref{eq2})
we find that the (absolute value of the) 
variation of $\alpha_3(M_Z)$ with respect to
$\sin^2\theta_{W}$ has a steeper behaviour for (\ref{eq1})
than for (\ref{eq2}) for values of $\sin^2\theta_{W}$ close to the
experimental point. Indeed,
\begin{equation}
\left|\frac{d \alpha_3^{-1}(M_Z)}{d (\sin^2\theta_{W})}
\right|_{MSSM}=\frac{15}{7}\frac{\sin^2\theta_{W}}{\alpha_{em}(M_Z)}< 
\left|\frac{d \alpha_3^{-1}(M_Z)}{d (\sin^2\theta_{W})}
\right|_{DMSSM}= 6\, \frac{\sin^2\theta_{W}}{\alpha_{em}(M_Z)}
\end{equation}
which means that the prediction for $\alpha_3(M_Z)$ close to the
 experimental point will vary 
faster in the DMSSM than in the MSSM, leading to potentially 
larger  two loop corrections in the DMSSM case. 

Finally,  a one-loop MSSM value  of $\alpha_3(M_Z)\approx 0.116$ 
differs significantly from its two loop value  $0.126 (\pm 0.01)$
(experimental value $0.119\pm 0.002$ \cite{pdg}) obtained 
using the unification  assumption. Thus two loop effects are 
important and an enhancement of such a difference is expected in the
DMSSM.
For the reasons outlined above we conclude that a careful analysis
of the DMSSM and of its gauge couplings running 
should include a full two loop RG approach.

\section{DMSSM: two-loop results}
A two-loop analysis of the DMSSM is easier than expected
since we  need perform only one-loop 
(wavefunction) calculations.
The result is indeed correct in two loop order for the gauge couplings.
For details see 
\cite{Shifman:1996iy},
\cite{Arkani-Hamed:2000mj}, \cite{Arkani-Hamed:1998ut}
with application to phenomenology in \cite{Ghilencea:1998mu}.
Further, the method has the advantage of unambiguously including
the threshold effects present at the scale $M_R$. This is
important because we  only need the bare value of $M_R$
for a two loop RG flow of the couplings \cite{Ghilencea:1998mu}
(we remind there are two Higgs pairs whose masses are
set equal to $M_R$ \cite{Aldazabal:2000sk} likely to 
bring threshold effects\footnote{In principle one  needs a string 
mechanism for fixing the moduli vev's giving the {\it bare} mass of 
these states.}). 
The renormalisation group (RG) evolution above the scale $M_R$ has the 
following structure 
\begin{equation} \label{rge1}
\alpha _{a}^{-1}(M_{R}) =-\Delta _{a}+\alpha_a^{-1}(M_U)
+\frac{B_{a}}{2\pi }\ln  \frac{M_U}{M_{R}} 
+\frac{3T_{a}(G)}{2\pi }\ln \left[ \frac{\alpha_a(M_U )}{\alpha 
_{a}(M_{R})}\right] ^{1/3}-\sum_{\phi }\frac{T_{a}(R_{\phi })}{2\pi }\ln \ 
Z_{\phi }(M_U ,M_{R})
\end{equation} 
where index ``a'' runs over indices 0,1$^{*}$,2,3 of the 
$U(1)_{B-L},\, SU(2)_R,\, SU(2)_L,\, SU(3)$ respectively\footnote{The
  index notation as 1$^*$ is chosen to distinguish it from that
  corresponding to $U(1)_Y$ below the scale $M_R$, see later.}. 
The values of the one-loop coefficients is given by
$B_a=\{3/2,3,3,-3\}_a$. The rather small value of 
$U(1)_{B-L}$ beta function in the  $k_{B-L}=32/3$ normalisation, 
will lead (for fixed value of the unified coupling) 
to a larger value of this coupling (in this normalisation)
at scale $M_R$ than in the   
$SU(2)$ groups case.
The quantities $\Delta_a$ may account for additional string
thresholds, which in our model are set to zero, as all complex
dimensions are twisted, thus are not expected to bring string corrections. 
 The coefficient $T_a(R_\phi)$ accounts for the Dynkin
index of the representation associated with the  gauge group ``a'' 
and its normalisation for ``a=0'' (as well as that of the
gauge couplings) in the notation of (\ref{rge1}) is that derived from
$\alpha_a(M_U)=\alpha$, for all indices ``a''.
The logarithm of the couplings accounts for pure
gauge effects (all orders) to the RG flow. The wavefunction 
coefficients $Z_{\phi}$ are
equal to unity at the tree level (i.e. one loop for the gauge couplings) thus 
account for two loop and beyond 
effects, induced by  the mixing matter-gauge or Yukawa effects.
In our analysis we only consider their one loop corrected
value (two loop for the gauge couplings running) induced by gauge effects
of $SU(3)\times SU(2)_L\times SU(2)_R\times U(1)_{B-L}$. The sum 
over $\phi$ runs over the entire matter spectrum and number of
generations. 
A simple one loop calculation of the coefficients $Z_{\phi}$ gives
\begin{equation}\label{zz}
Z_{\phi}(M_U,M_R)=\prod_{a=0}^{3}
\left[\frac{\alpha_a{M_U}}{\alpha_a{M_R}}\right]^{-2 C_a(\phi)/B_a}
\end{equation}
where we used the notation 
$C_a(\phi)~=~\{ Q^2_{B-L}/k_{B-L},3/4,3/4,4/3\}_a$
for the quadratic Casimir operator of
$U(1)_{B-L}$, $SU(2)_R$, $SU(2)_L$ and  $SU(3)$ respectively. 
Additional effects on coefficients $Z_\phi$ 
are expected from Yukawa interactions, 
controlled by the superpotential terms and 
strongly model dependent in this case.
For this reason we do not include them; they may be 
accounted for by using in (\ref{rge1}) the replacement 
$Z_\phi\rightarrow Z_\phi \times Z_\phi^y$ with
${d}/{dt} \ln Z_{\phi}^y(M_U,M_R) =
\sum_{\nu}A_\nu(\phi)y_\nu(t),\,
t=1/(2\pi)\ln(scale)
$ 
accounting for  Yukawa one loop wavefunction renormalisation 
and coefficients $A_\nu$ depending on the superpotential.
This relation may be integrated analytically to give 
Yukawa correction $Z_\phi^y$ to eq.(\ref{zz}).

The RG flow below the scale $M_R$ is that familiar for the MSSM with
the important observation that the hypercharge normalisation 
which affects  one and two loop contributions is that 
corresponding to 3/11 (and not 3/5 as in the MSSM).  The 
equations have the structure
\begin{equation} \label{rge2}
\alpha _{i}^{-1}(M_{Z}) =-\delta _{i}+\alpha_i^{-1}(M_R)
+\frac{b_{i}}{2\pi }\ln  \frac{M_R}{M_{Z}} 
+\frac{3T_{i}(G)}{2\pi }\ln \left[ \frac{\alpha_a(M_R )}{\alpha 
_{a}(M_{Z})}\right] ^{1/3}-\sum_{\psi }\frac{T_{i}(R_{\psi })}{2\pi }\ln \ 
Z_{\psi }(M_R ,M_{Z})  \label{rge_general} 
\end{equation} 
with index ``i'' running over 1,2,3 associated with 
the groups $U(1)_{Y}$, $SU(2)_L$ and
$SU(3)$ respectively. One loop coefficients  $b_i=\{3,1,-3\}$ and
 the coefficients $Z_{\psi}$ are similar to those in eq.(\ref{zz}) 
\begin{equation}\label{zzz}
Z_{\psi}(M_R,M_Z)=\prod_{i=1}^{3}
\left[\frac{\alpha_i{M_R}}{\alpha_i{M_Z}}\right]^{-2 C_i(\psi)/b_i}
\end{equation}
with $C_i(\psi)~=~\{Q^2_{Y}/k_{Y},3/4,4/3\}_i$.
To include the Yukawa effects below the scale $M_R$, 
an approach similar to that above $M_R$  may be used.
\begin{figure}[t] 
\begin{tabular}{cc|cr|} 
\parbox{8cm}{ 
\psfig{figure=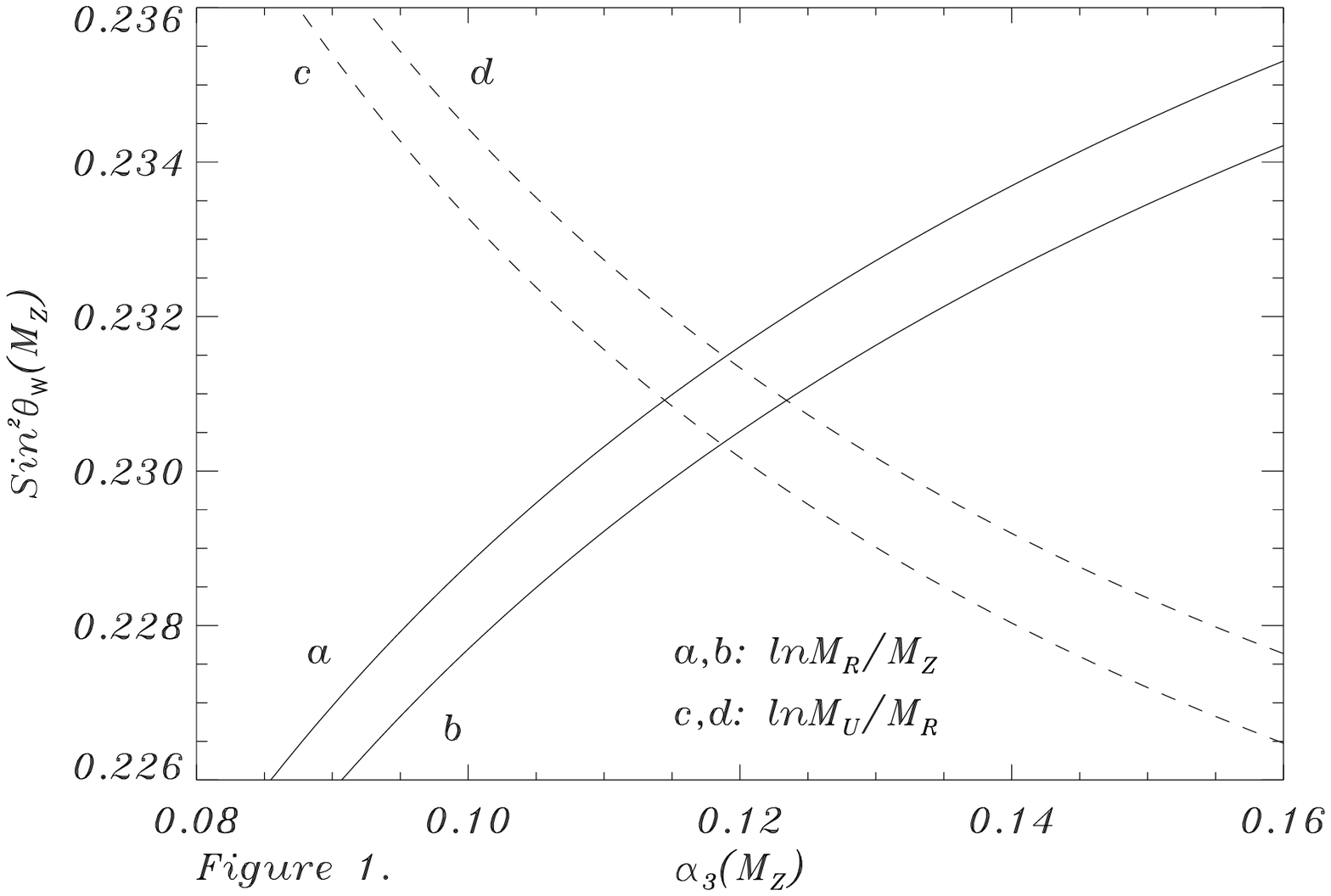,height=7.5cm,width=7.8cm}} 
\hfill{\,\,\,\,\,\,\,\,\,\,\,} 
\parbox{8cm}{ 
\psfig{figure=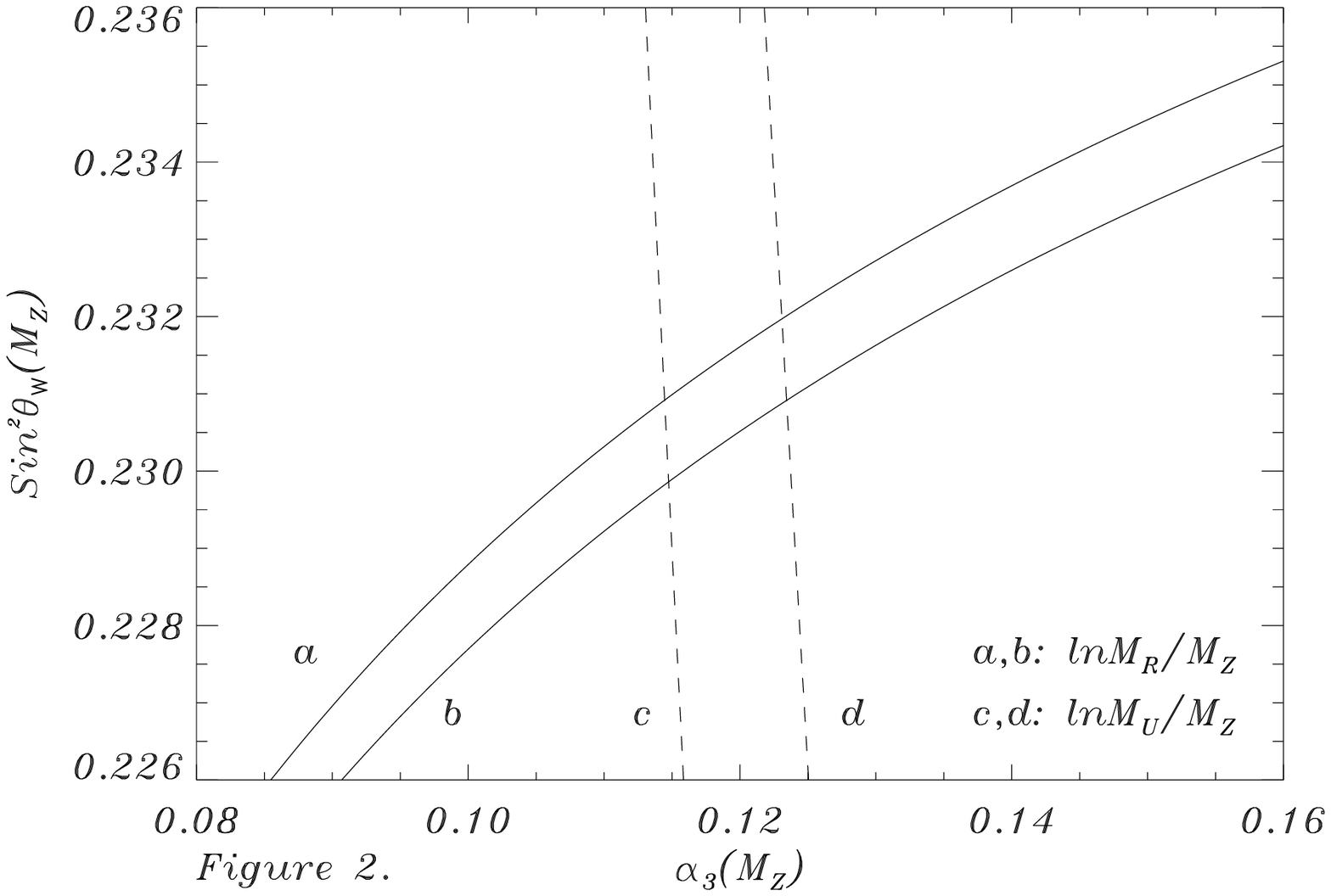,height=7.5cm,width=7.8cm}}  
\end{tabular} 
\newline 
\newline \newline 
{\small Figure 1:
Curves: a: $\ln M_R/M_Z=4.05$;  b: $\ln M_R/M_Z=4.40$;
c: $\ln M_U/M_R=18.5$;  d: $\ln M_U/M_R=18.9$.  
The curves mark the limiting values
(conservative estimates) of  $M_R$ and $M_U$ for which 
one may still simultaneously fit the experimental constraints on 
$\alpha_3(M_Z)$ and $\sin^2\theta_W(M_Z)$.\\
\noindent
Figure 2: As for Figure 1 with curves: 
a: $\ln M_R/M_Z=4.05$; b: $\ln M_R/M_Z=4.40$;
c: $\ln M_U/M_Z=22.55$; d: $\ln M_U/M_Z=23.30$. \\}
\end{figure} 

A numerical investigation of the above equations 
gives the results of Figures 1-3. Provided that unification 
takes place,  the following (conservative) 
two-loop bounds may be placed on the
intermediate scale $M_R$: $M_Z \times exp(4.05)\leq M_R\leq M_Z \times
exp(4.4)$ and on the unification 
scale $M_Z \times exp(22.55)\leq M_U\leq M_Z \times exp(23.30)$
which give: $5.233 \,\,TeV\leq M_R\leq 7.427\,\, TeV$ and 
$5.665\times 10^{11}\,GeV \leq M_U\leq 1.199 \times 10^{12}\, GeV$.
We notice a rather strong variation of the predicted 
two loop value of $M_R$ from
its one loop value of about 1 $TeV$ \cite{Aldazabal:2000sk}
compatible with the same low energy input for
$\alpha_3(M_Z)$ and $\sin^2\theta_{W}(M_Z)$.
This is essentially due to additional radiative effects induced by the 
larger (than in the MSSM) (non-Abelian) gauge group. These effects are
further strengthened by the presence of the  logarithm
in front of $M_R$ in the RG equations (\ref{rge1}),(\ref{rge2}).

Similar to the MSSM case,  our results,  Figures 1,2,3
are sensitive to the value of $T_{eff}$, eq.(\ref{rge1})  and which 
has so far been taken equal to $M_Z$.
$T_{eff}$ is a function of the low 
energy supersymmetric thresholds $\delta_i$ which affect the
prediction of the correlation $\alpha_3 - \sin^2\theta_{W}$
at $M_Z$ \cite{Langacker:1993rq}. $T_{eff}$
is changed from the MSSM case due to the non-standard hypercharge
normalisation and different RG flow above the scale $M_R$.
It parametrises our lack of detailed knowledge of the low energy
supersymmetry spectrum
(similar to the MSSM case) as an overall effect on $\alpha_3(M_Z)$.
As a result, its value in eq.(\ref{eq1}) is a 
combination of $\delta_i$'s of (\ref{rge2}) giving
\begin{equation}\label{teff}
T_{eff}=M_Z \frac{M^{\frac{3}{10}}_{\tilde L} 
M^{\frac{3}{2}}_{\tilde Q}} {M^{\frac{6}{5}}_{Z} 
M^{\frac{3}{10}}_{\tilde U} M^{\frac{3}{10}}_{\tilde E}}
\left[\frac{\mu}{M_Z}\right]^{\frac{2}{15}}
\left[\frac{M_2}{M_Z}\right]^{\frac{2}{5}}
\left[\frac{M_3}{M_Z}\right]^{\frac{2}{15}}
\left[\frac{M_H}{M_Z}\right]^{\frac{1}{30}}
\approx M_Z
\left[\frac{M_{\tilde L}}{M_Z}\right]^{\frac{6}{5}}
\left[\frac{M_{2}}{M_Z}\right]^{\frac{2}{5}}
\left[\frac{M_{3}}{M_Z}\right]^{\frac{2}{15}}
\end{equation}
where the approximation only holds for degenerate squarks and
sleptons, hence $T_{eff}$ increases with their 
mass and also with that of gauginos.
\begin{figure}[ht] 
\label{mssmfigure0}  
\begin{tabular}{cc|cr|} 
\parbox{8cm}{ 
\psfig{figure=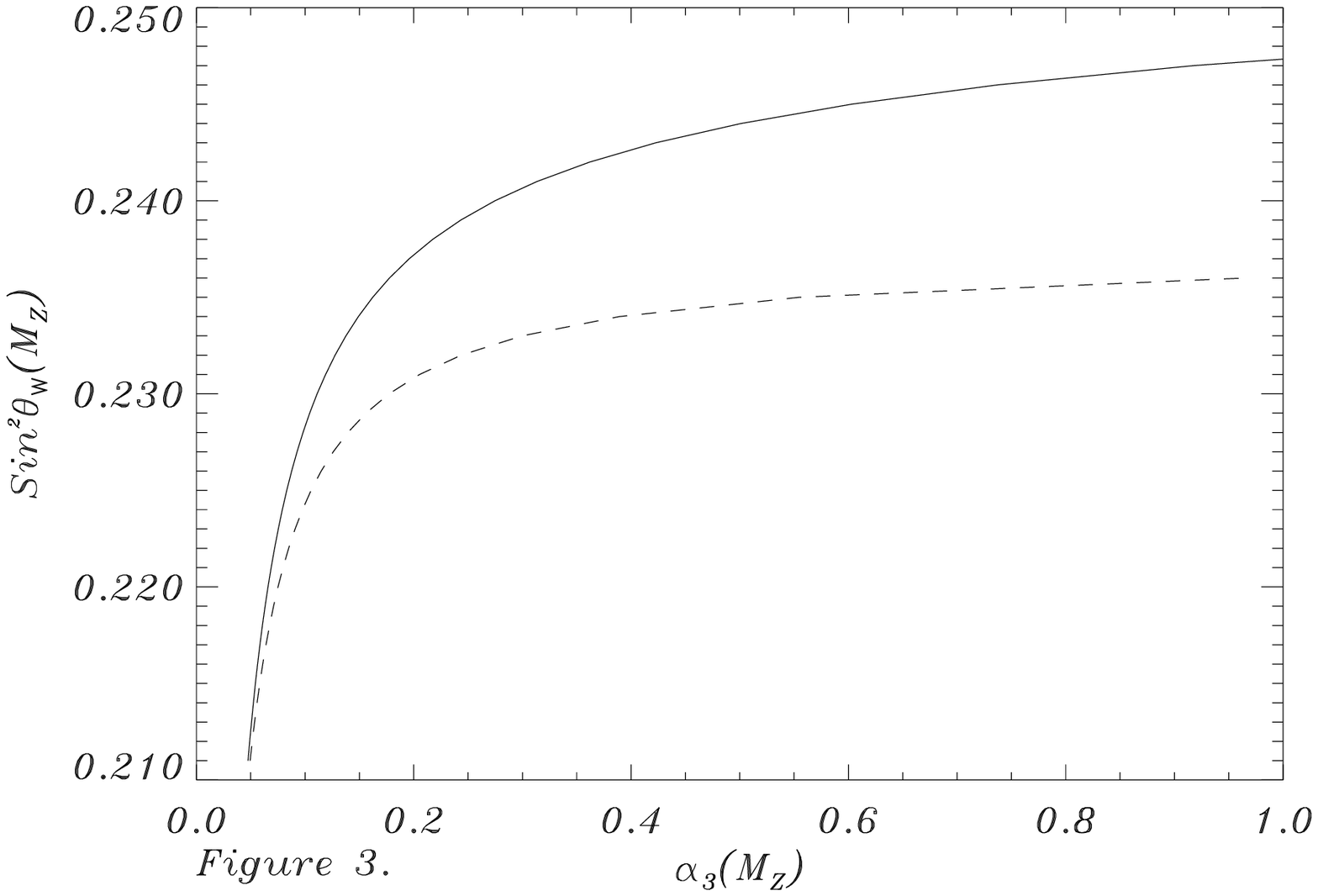,height=7.5cm,width=7.8cm}} 
\hfill{\,\,\,\,\,\,\,\,\,\,\,} 
\parbox{8cm}{ 
\psfig{figure=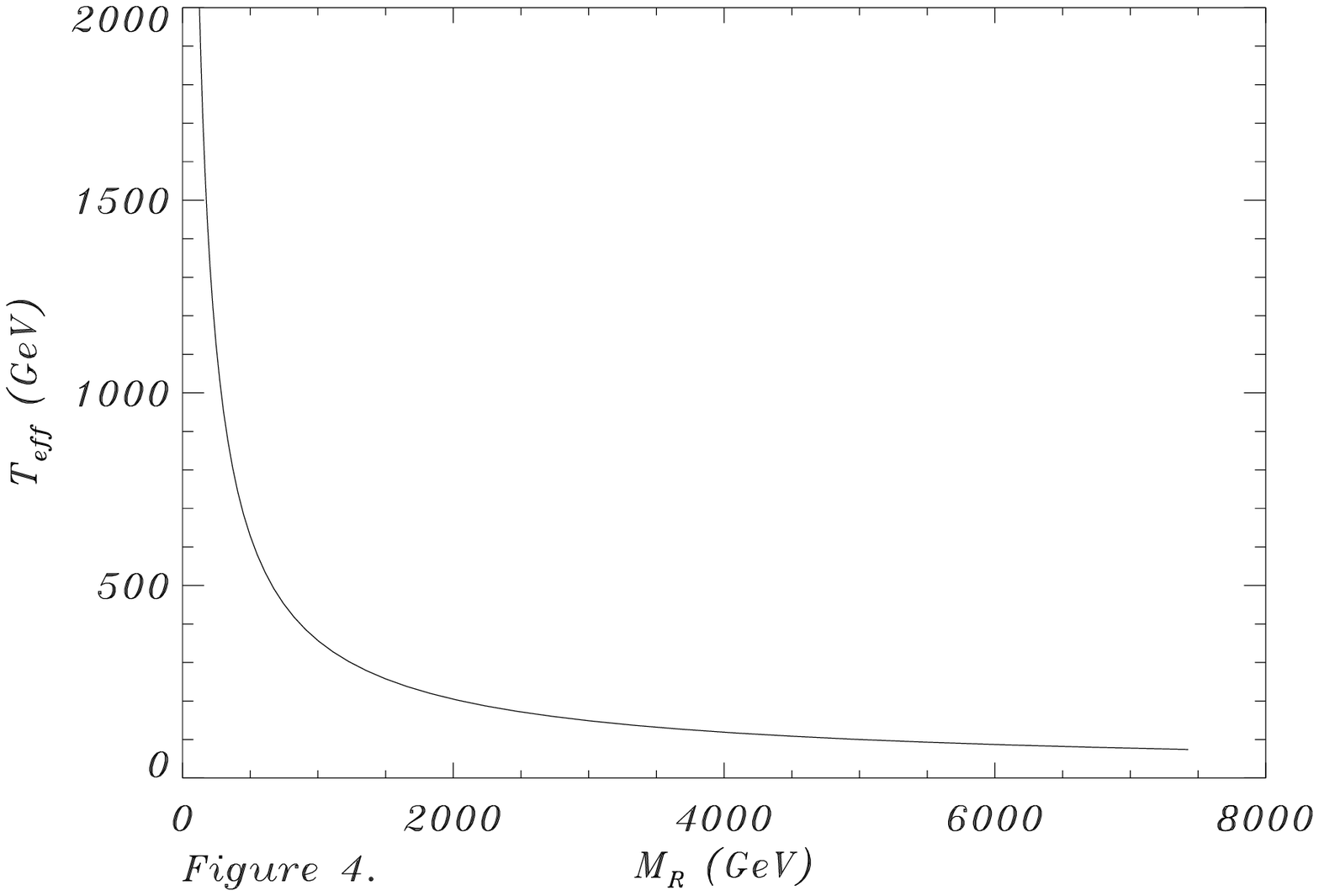,height=7.5cm,width=7.8cm}}  
\end{tabular} 
\newline 
\newline 
\newline 
{\small 
Figure 3. Two-loop (continuous line) correlation for $\ln M_R/M_Z=4.2$
($M_R\approx 6 TeV$). The one-loop case (dashed line) corresponding to 
the same value of $M_R$ cannot fit the low energy values of
$\alpha_3(M_Z)$, $\sin^2\theta_{W}(M_Z)$ unless  $M_R$ is reduced 
by a factor of $\approx 4$, in 
the region of 1.5 TeV. This shows a significant change of  
the one-loop from the two-loop prediction for $M_R$
with $\alpha_3(M_Z)$, $\sin^2\theta_W(M_Z)$ fixed to the 
experimental values.\\
\noindent
Figure 4. Two-loop plot of $T_{eff}$ versus $M_R$ for
$\alpha _{3}(M_{Z})$ and $\sin^2 \theta_{W}(M_Z)$
fixed to their (central) experimental values.}
\end{figure} 
\begin{figure}[ht]
\begin{tabular}{cc|cr|} 
\parbox{8cm}{ 
\psfig{figure=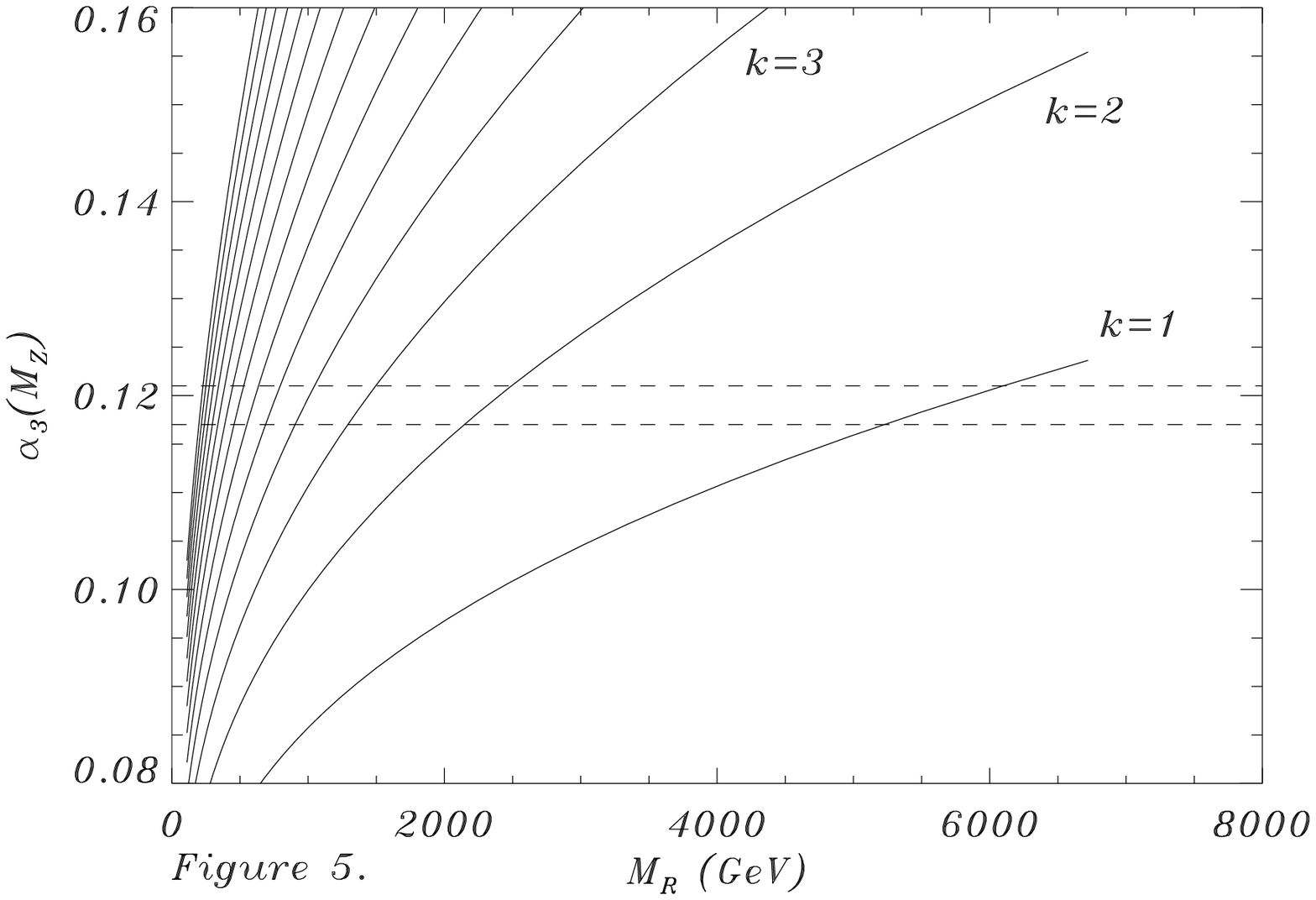,height=7.5cm,width=7.8cm}} 
\hfill{\,\,\,\,\,\,\,\,\,\,\,} 
\parbox{8cm}{ 
\psfig{figure=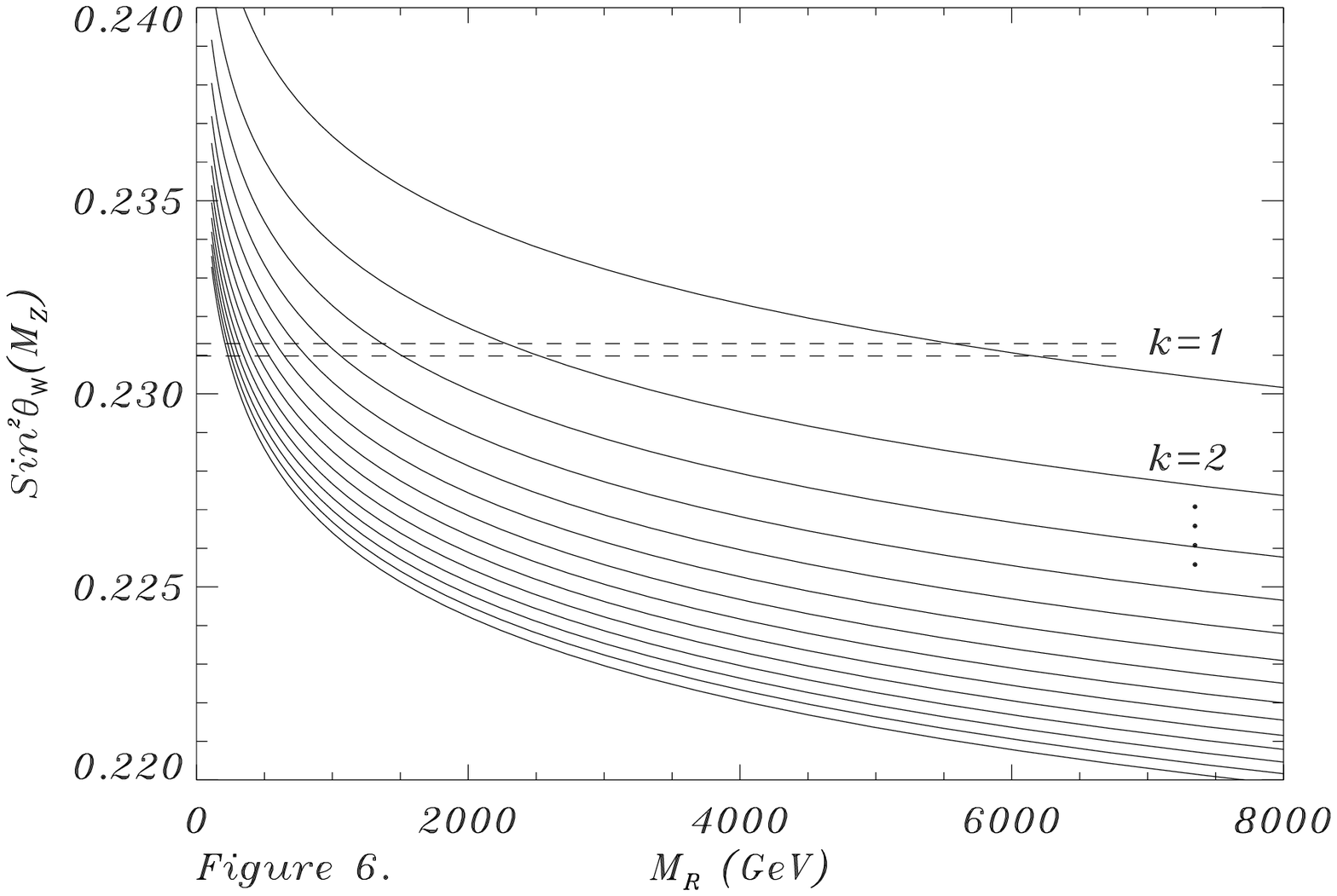,height=7.5cm,width=7.8cm}}  
\end{tabular} 
\newline
\newline
\newline
{\small 
Figure 5. Parametric 
plot of $\alpha_{3}(M_{Z})$ versus $M_R$ ($\sin^2 \theta_{W}$
fixed to its experimental range), computed for different values 
of the low energy supersymmetric threshold $T_{eff}=k M_Z$ with $k$
increasing from right to  left by unity, as shown in the 
figure. The dashed lines mark the experimental limits.\\
Figure 6.  Parametric plot for $\sin^2\theta_{W}$ versus  $M_R$ 
($\alpha_3(M_Z)$ fixed to $0.119$),  calculated for different values 
of the low energy supersymmetric threshold $T_{eff}=k M_Z$ with $k$
increasing downwards, as shown in the 
figure. The dashed lines mark the experimental
limits.}
\end{figure} 

At one-loop level the effect of increasing  $T_{eff}$ is to increase
$\alpha_3(M_Z)$, see eq.(\ref{eq1}). 
This behaviour is different from the MSSM case where the
opposite effect is manifest (see eq.(\ref{eq2}) with
$T_{eff}\rightarrow M_{eff}$). The difference is in essence 
due to different RG flow/gauge structure above the scale $M_R$.
The increasing effect on $\alpha_3(M_Z)$ due to (increasing) $T_{eff}$  
may be compensated for by decreasing $M_R$ (see Figures~1,4) 
and also $M_U$ (see Figure~2). 
This may also be seen in one loop order from
eqs.(\ref{eq1}), (\ref{teff}) which show that
for fixed low energy input,  the combination
$M_R \times T_{eff}$ must stay constant in this approximation.
This is an important effect that we would like to stress. 
We mentioned that pure one loop effects predict
($\alpha_3(M_Z)$, $\sin^2\theta_{W}$ fixed) a  value for $M_R$ of order 
$1\,\, TeV$ which is further increased by two
loop effects to values of order $5-7\, TeV$. 
This may be a concern since one must explain why the mass
of the additional Higgs sector (also of mass equal to $M_R$)
is so high compared to the electroweak scale. However, low energy 
supersymmetric thresholds ($T_{eff}$) can reduce significantly 
the predicted value 
of $M_R$ for fixed low energy input, see Figure~4.
For example  sleptons and gaugino masses with (lower bounds on) masses 
in the region of 300 (200) GeV  increase the value of
$T_{eff}$ (initially set to $M_Z$) 
by a factor of $\approx 7.90 (3.91)$ which thus reduces $M_R$
(Figure~4) in the region 450 (1000) GeV. For a 
general picture of this situation see also Figures 5 and 6
where the dependence of $\alpha_3$ and $\sin^2\theta_W$ is 
showed as a parametric plot for different values of $T_{eff}$ 
and $M_R$. Therefore, the
presence of a low energy supersymmetric spectrum 
ensures that there is a left-right symmetry breaking scale
not far above $M_Z$. This avoids  the difficulty and 
need for a rather large $M_R$ of Figures 1,2,3  where $T_{eff}=M_Z$.
In addition, (lower) bounds on $M_R$  may also exist to avoid FCNC problems
rather generic \cite{Ma:1995ge} in ``left-right'' symmetric models
with $M_R$ in the region of 1 $TeV$. This issue was addressed in 
\cite{Ma:1995ge} and it would be useful to have a 
detailed investigation in the DMSSM model. The (lower) 
bounds on $M_R$ would in turn be related to $T_{eff}$ (see Figure~4) to 
provide  upper bounds for it and thus for 
(the combination of) the low energy supersymmetric spectrum 
(via eq.(\ref{teff})).

\section{Conclusions and outlook}
We presented a simple  method to compute two-loop effects in a model
with different symmetry groups above/below the scale $M_R$
and to account for the threshold effect at this scale.
The model is successful in  achieving a low scale of unification with 
logarithmic running only for the gauge couplings. The low scale of 
unification is due to the enhanced gauge symmetry and non-standard
hypercharge $U(1)_Y$ and $U(1)_{B-L}$ normalisations. 
We showed that the two loop  prediction for the ``left-right'' symmetry 
scale is the result of the competing effects between pure two loop terms
and one loop supersymmetric thresholds. The latter thus ensure that 
the value of $M_R$ may be kept rather small, of order  of 1 $TeV$ or
even less. Further quantitative analysis of this  model is however 
required to  explain how the breaking of the ``left-right'' 
symmetry is induced.  Finally, a comparative analysis to 
the MSSM case of the correlation
($\alpha_3(M_Z), \sin^2\theta_W(M_Z))$ (as in \cite{Ghilencea:2001qq})
could  provide an insight into
the relative amount of fine tuning of the high scale 
one needs perform to keep such correlation stable against high scale 
physics. Such analysis would help understand which model, DMSSM or
MSSM is more predictive/less fine-tuned.
\\

\noindent 
{\bf Acknowledgements}\\

\noindent
The author would like to thank Fernando Quevedo and Louis Ib\'a\~nez 
for helpful  discussions on this work.
The work was supported by the  University of Bonn under the 
European Commission RTN programme HPRN-CT-2000-00131.

\end{document}